\documentclass[aip,apl,twocolumn,showpacs,10pt]{revtex4-1}

\usepackage{graphicx} 
\usepackage{color}
\definecolor{Green}{rgb}{0,0.5,0}

\usepackage[english]{babel}
\usepackage[utf8]{inputenc}

\usepackage{amssymb}

\usepackage{ulem}

\usepackage{textcomp}

\begin{document}

\title{Graphene nanoribbons with wings}
\author{D. Bischoff}
\email{dominikb@phys.ethz.ch}
\affiliation{Solid State Physics Laboratory, ETH Zurich, CH-8093 Zurich, Switzerland}
\author{M. Eich}
\affiliation{Solid State Physics Laboratory, ETH Zurich, CH-8093 Zurich, Switzerland}
\author{F. Libisch}
\affiliation{Institute for Theoretical Physics, Vienna University of Technology, A-1040 Vienna, Austria}
\author{T. Ihn}
\affiliation{Solid State Physics Laboratory, ETH Zurich, CH-8093 Zurich, Switzerland}
\author{K. Ensslin}
\affiliation{Solid State Physics Laboratory, ETH Zurich, CH-8093 Zurich, Switzerland}
\date{\today}



\begin{abstract}
We have investigated electronic transport in graphene nanoribbon devices with additional bar-shaped extensions (``wings'') at each side of the device. We find that the Coulomb-blockade dominated transport found in conventional ribbons is strongly modified by the presence of the extensions. States localized far away from the central ribbon contribute significantly to transport. We discuss these findings within the picture of multiple coupled quantum dots. Finally, we compare the experimental results with tight-binding simulations which reproduce the experiment both qualitatively and quantitatively.
\end{abstract}



\maketitle


Graphene~\cite{Novoselov2004} -- a monolayer of carbon atoms -- possesses a variety of novel electronic properties including a linear energy dispersion relation.~\cite{Geim2007} Nanodevices  made from graphene were recognized as interesting and potentially useful building blocks for applications and quantum circuits.~\cite{Trauzettel2007,Schwierz2010} Narrow graphene stripes called nanoribbons~\cite{Chen2007,Han2007} are the basic building blocks of more complicated graphene nanoelectronic devices. If they have perfect edges, they are expected to either exhibit a well defined band gap or conducting edge states depending on the edge orientation.~\cite{Nakada1996}

Experimentally, most graphene nanoribbons show a region of strongly suppressed conductance close to the charge neutrality point.~\cite{Oezyilmaz2007,Han2007,Molitor2009-PRB,Bischoff2012} At sufficiently low temperature, Coulomb blockade is typically observed in this regime of suppressed conductance suggesting that localized charge carriers dominate the electronic transport properties.~\cite{Oezyilmaz2007,Molitor2009-PRB,Bischoff2012} Various mechanisms were suggested to be responsible for the observed localization of charge carriers.~\cite{Chen2007,Sols2007,Adam2008-PRL,Stampfer2009-PRL,Oostinga2010,Han2010-Ribbons,Droescher2011} Recent work has shown that for sufficiently clean devices, edge disorder is mainly responsible for the experimentally observed suppressed conductance~\cite{Bischoff2012,Smith2013} and that the sites of localized charge can be close to the ribbon edges.~\cite{Todd2009,Stampfer2009-PRL,Bischoff2015,Simonet2015} In a recent experiment, evidence was found, that certain electronic states are mainly localized along the disordered graphene edge and can extend along the device edge into the wider graphene leads.~\cite{Bischoff2014}

\begin{figure}[tbp]
	\centering
	\includegraphics[width=0.98\columnwidth]{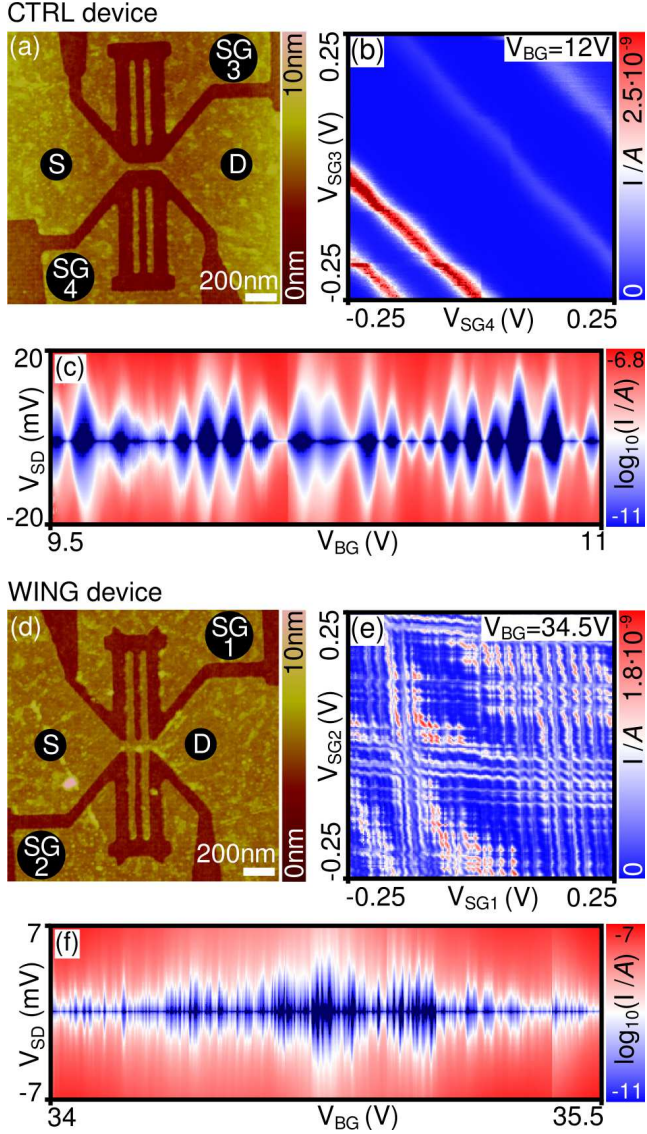}
	\caption{(a) Scanning force microscopy (SFM) image of one of the control (CTRL) devices. In the brown parts, the graphene (yellow) was etched away. The different contacts are marked by black circles. The additional uncontacted graphene stripes above and below the ribbon were patterned in order to keep the device geometry similar to the WING devices. (b) Current flowing between S and D in one CTRL device as a function of the applied side gate voltages ($V_\mathrm{SD}=1\,\mathrm{mV}$, $T=4.2\,\mathrm{K}$). (c) Current flowing through one CTRL device as a function of the applied back gate voltage and source-drain bias ($T=4.2\,\mathrm{K}$, $V_\mathrm{SG3}=V_\mathrm{SG4}=0\,\mathrm{V}$). (d) SFM image of one of the ribbons with wings (WING): additional graphene stripes were patterned at the side of the main graphene nanoribbon. (e) Current flowing through the WING device as a function of the applied side gate (SG) voltages ($T=1.6\,\mathrm{K}$, $V_\mathrm{SD}=1\,\mathrm{mV}$). (f) Current flowing through the WING device as a function of the applied BG and bias voltages ($T=1.6\,\mathrm{K}$, $V_\mathrm{SG1}=V_\mathrm{SG2}=0\,\mathrm{V}$). For the measurements in (b,e), the same ranges in SG voltage were applied. For (c,f), the same range in BG voltage was applied, whereas a different range in bias voltage was used. In (c,f), the current is plotted on a logarithmic scale.}
	\label{Figure1}
\end{figure}

In this paper we investigate a device geometry which we call ``nanoribbons with wings'', as the device resembles the outline of a dragonfly. The investigated devices consist of a typical nanoribbon design where additional graphene stripes of similar width are attached perpendicularly at each nanoribbon edge (see Fig.~\ref{Figure1}d). This device geometry was designed to study the interplay between transport along the edges and through the bulk of the ribbon devices, as the total edge length is significantly increased while the device area connecting source and drain contacts is maintained. We show that the additional wings significantly alter the electronic transport properties of the ribbons. Transport in these winged devices is governed by charge localization leading to Coulomb blockade. Notably, localized states sitting far out in the wings and therefore far away from the direct current path contribute to transport. The typical area on which charges are localized is estimated using measurements in magnetic field. The findings are finally compared to tight-binding simulations.

The two investigated device geometries are shown in Figs.~\ref{Figure1}a,d. The difference between the nanoribbon control devices (CTRL) shown in Fig.~\ref{Figure1}a and the nanoribbons with  wings (WING) in Fig.~\ref{Figure1}d are the extra graphene bars added at each side to the latter ones. The two investigated CTRL graphene nanoribbons are about $220\,\mathrm{nm}$ long and $50\,\mathrm{nm}$ wide. The central ribbon of the three investigated WING devices is of similar length and between $50-70\,\mathrm{nm}$ wide. The added wings are about $50\,\mathrm{nm}$ wide and $550\,\mathrm{nm}$ long. All devices were characterized in at least two different cooldowns and at different applied gate voltages. While details vary, the general findings presented in this paper are valid for all investigated devices and cooldowns.

The mechanically exfoliated~\cite{Novoselov2004} single layer graphene flakes~\cite{Ferrari2006,Graf2007} were  patterned by reactive ion etching.~\cite{Molitor2009-PRB,Bischoff2012} The graphene nanodevices are situated on top of a $285\,\mathrm{nm}$ thick SiO$_2$ substrate with a global silicon back gate (BG). For each device, DC voltages were applied between source/drain, at each side gate (SG) and at the BG. Current between source and drain was recorded as a function of the applied voltages. No leakage currents could be experimentally detected from any gate to any other part of the device. Measurements were performed at temperatures of $T=1.6\,\mathrm{K}$ and at $T=4.2\,\mathrm{K}$.

Both CTRL and WING devices were initially characterized by recording the current as a function of the applied BG voltage at a small constant bias. In agreement with previous experiments, a region of suppressed conductance is found.~\cite{Oezyilmaz2007,Han2007,Molitor2009-PRB,Bischoff2012} This region is typically at positive BG voltages due to unintentional chemical doping of the devices. The  measurements discussed in this paper are all recorded in this regime of suppressed conductance.

Transport through the CTRL devices is governed by Coulomb blockade (CB) in agreement with previous studies.~\cite{Oezyilmaz2007,Molitor2009-PRB,Bischoff2012} When measuring the current as a function of the applied BG and bias voltage at constant applied SG voltages, CB diamonds are observed as shown in Fig.~\ref{Figure1}c. Further information about the position of the localized charges resulting in CB is obtained by following the position of a CB peak while changing the applied gate voltages.~\cite{Wiel2003,Bischoff2013} The current flowing through one CTRL device as a function of applied SG voltages is shown in Fig.~\ref{Figure1}b: diagonal lines with a slope of minus one indicate that the localized states couple equally well to both SGs. This is not surprising as the two symmetric SGs are located far away from the ribbon and therefore cannot discriminate potentially existing localized states at different positions inside the ribbon.~\cite{Todd2009,Stampfer2009-PRL,Bischoff2015,Simonet2015,Pascher2012}

Transport in the WING devices is also governed by CB as shown in Fig.~\ref{Figure1}f. However, the diamonds are about a factor of 10 narrower in BG and a factor of 4 smaller in bias direction. Also, the evolution of CB peaks in SG voltages is clearly different as shown in Fig.~\ref{Figure1}e: many closely spaced and oscillating lines are found running nearly vertically or horizontally through the measurement. Additionally, a modulation of the current amplitude is found when varying the SG voltages.

The nearly vertical/horizontal resonances observed in the WING devices show that areas much closer to one SG than to the other have an influence on transport. This would, for example, not be the case in a diffusive and incoherent metallic system. In one of the WING devices, the outermost $200\,\mathrm{nm}$ of the wings were removed by RIE after the device was characterized. This new device with shortened wings exhibited qualitatively and quantitatively similar transport properties to the unshortened version. The only notable exception is that the slopes of the nearly vertical/horizontal lines got flatter/steeper, indicating that the outermost $200\,\mathrm{nm}$ of the wings did contribute to the transport characteristics.

\begin{figure}[tbp]
	\includegraphics[width=\columnwidth]{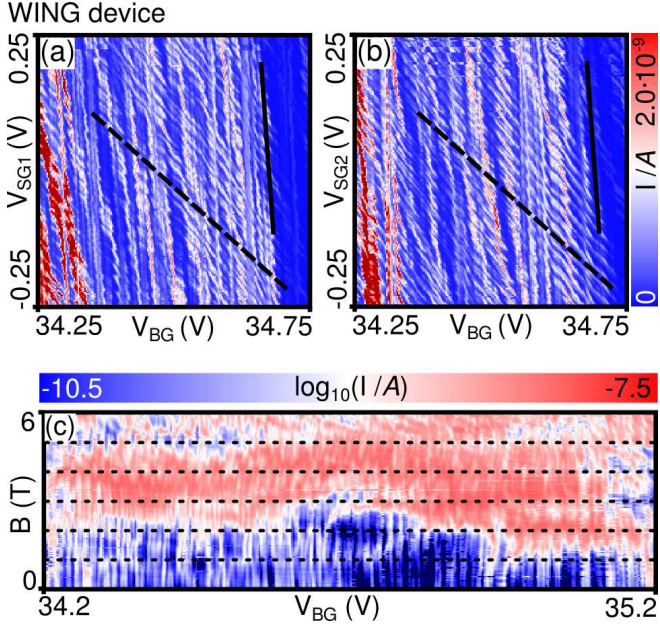}
	\caption{(a,b) Current flowing through one WING device as a function of one SG and the BG voltage. The other SG is kept at ground potential ($T=1.6\,\mathrm{K}$, $V_\mathrm{SD}=1\,\mathrm{mV}$). While the overall behavior is similar, the fine features are distinctly different. As a guide to the eye, two features with different slopes are marked with black  lines. (c) Current flowing through one WING device as a function of the applied BG voltage and the perpendicular magnetic field ($T=1.6\,\mathrm{K}$, $V_\mathrm{SG1}=V_\mathrm{SG2}=0\,\mathrm{V}$, $V_\mathrm{SD}=1\,\mathrm{mV}$). The current stays unchanged up to fields of $1-2\,\mathrm{T}$. At fields of about $3\,\mathrm{T}$, the current increases by roughly one order of magnitude (note logarithmic colorscale).}
	\label{Figure2}
\end{figure}

Further information is obtained by measuring the current as a function of the applied BG and SG voltages as shown in Figs.~\ref{Figure2}a,b: both plots look very similar at a first glance, despite the fact that in one case SG1 (a) and in the other case SG2 (b) was used. This indicates that the part of the device being equally strongly coupled to both SGs, i.e. the central ribbon part, determines the overall conductance. The finer lines with flatter slopes (indicating stronger coupling to the SGs) differ for the two measurements in Figs.~\ref{Figure2}a,b, indicating that the transport details are also influenced by the wings.

Next, the spatial extent of localized states is investigated. Fig.~\ref{Figure2}c shows transport data of a WING device obtained in a perpendicular magnetic field. Such plots were also recorded as a function of the SG voltages and show similar behavior. Up to magnetic fields of about $1\,\mathrm{T}$, the resonances remain mostly unchanged. In the region of $1-2\,\mathrm{T}$, the CB resonances start to move, broaden and increase in amplitude. At higher fields, starting from $2-3\,\mathrm{T}$, the current suddenly increases by more than an order of magnitude. Assuming that the addition of one flux quantum to the area of a localized state will significantly change transport~\cite{Fock1928,Darwin1930,Beenakker1991}, the following areas can be extracted: $A(1\,\mathrm{T})=h/(eB) \approx (65\,\mathrm{nm})^2$ and $A(2\,\mathrm{T})\approx (45\,\mathrm{nm})^2$. The strong increase in current above $2-3\,\mathrm{T}$ can be attributed to reduced backscattering in electronic transport due to the chirality induced by the magnetic field.~\cite{Prange1988} Alternatively, one could try to estimate the area of a localized state from the width of a CB diamond.~\cite{Wiel2003} This approach is however not useful here as it is experimentally challenging to determine which resonance belongs to which localized state.



\begin{figure}[tbp]
	\includegraphics[width=\columnwidth]{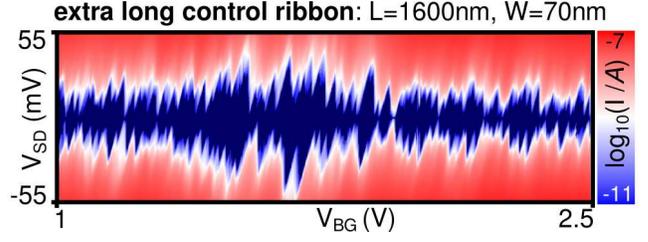}
	\caption{Current flowing through a $1.6\,\mathrm{\mu m}$ long CTRL-type device as a function of the applied BG and bias voltage ($T=4.2\,\mathrm{K}$, no side gates). The size of the CB diamonds is about an order of magnitude larger than for the WING devices.}
	\label{Figure3}
\end{figure}

In summary, the experimental findings suggest a picture of tunneling coupled states of varying spatial extent whose locations are spread throughout the whole structure including the remote ends of the wings. These states interact via Coulomb interaction~\cite{Hofmann1995,Baines2012}, resulting in a huge multi-quantum-dot molecule. Due to the overlap of the wave-functions, current flow through the structure can also be influenced by states localized far away from the central ribbon. The enhanced number of Coulomb-blockade diamonds in Fig.~\ref{Figure1}f is the manifestation of an enhaced number of states per energy window and ribbon length due to the device area being increased by the wings. In such a situation, the distinction between current flowing along the edge or within the bulk becomes meaningless. This contrasts with a CTRL-type ribbon of increased length shown in Fig.~\ref{Figure3}, where the area is increased but the number of states per energy window and ribbon length remains the same. As a result, the diamonds overlap strongly due to the serial arrangement of the localized states.


\color{black}

\begin{figure}[tbp]
	\includegraphics[width=\columnwidth]{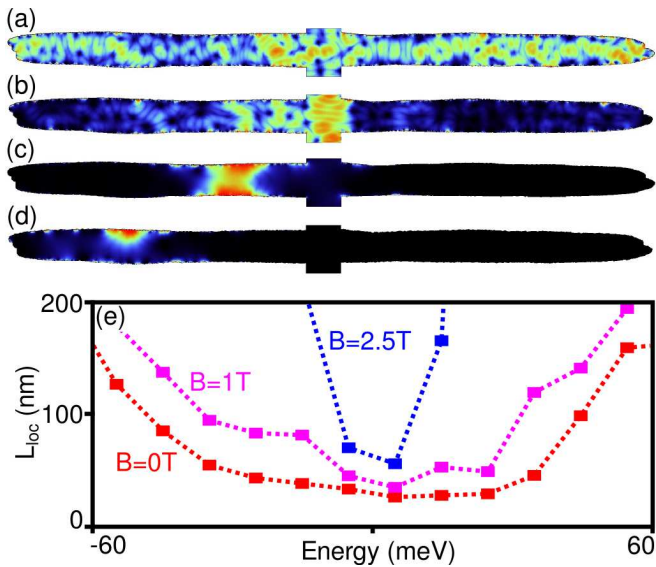}
	\caption{Envelopes of states simulated for the geometry of a central ribbon part with one set of attached wings. (a) State mostly delocalized. (b) State coupling directly through the middle ribbon element. (c) ``Square'' state coupling both ribbon edges together. (d) A state mostly localized along one ribbon edge. (e) Averaged localization length $L_\mathrm{loc}$ of scattering states along the wing as function of energy [see Eq.~(\ref{eq:decay})], for three different magnetic field strengths. Each square represents an average over 80 scattering states.}  
	\label{Figure4}
\end{figure}

In order to confirm this general picture and to obtain more insights into the spatial distribution of wave functions, we have performed tight-binding simulations using third-order nearest neighbor tight binding.~\cite{Libisch2009} For an analysis of localized eigenstates, we consider a central ribbon element with one set of wings of experimental dimensions (see Fig.~\ref{Figure4}). We connect the top and bottom of the center region using periodic boundary conditions to obtain a simulation cell of manageable size. We assume no bulk disorder and a randomly jagged edge with an edge roughness of the order of $2\,\mathrm{nm}$ (with $5\,\mathrm{nm}$ correlation length). We identify four different types of wave-functions as shown in Figs.~\ref{Figure4}a-d: (a) states mostly delocalized over the whole structure; (b) states mostly located in the central ribbon; (c) ``square'' states localized in the wings, coupling both edges; and (d) states localized predominantly along one edge of the wings.  The delocalized states (a) are mostly found far away from the Dirac point. When approaching the Dirac point, fewer states couple  directly through the central ribbon (b). Instead, states localize either at a single edge (d), or between the two edges of the wing (c). 

To extract length scales from the simulations, we calculated scattering states inside a wing geometry. We find exponential (Anderson) localization in direction $x$ along the wing ~\cite{Libisch2012}
\begin{equation}\label{eq:decay}
\left|\psi(x)\right|^2 = \int \left|\psi(x,y)\right|^2\,dy  \approx A \times \exp(-x / L_{\mathrm{loc}}).
\end{equation}
The localization length $L_\mathrm{loc}(\varepsilon)$ as a function of energy is extracted by averaging over 80 scattering states in a narrow energy window ($\varepsilon \pm 5\,\mathrm{meV}$). $L_{\mathrm{loc}}(\varepsilon)$ decreases close to the Dirac point (see red curve in Fig.~\ref{Figure4}e). A localization length of the order of $50\,\mathrm{nm}$ (the ribbon width) is consistent with experiment. States located in either flap contribute to Coulomb blockade peaks, where the overlap of their exponential tails with the center region of the dot is responsible for the tunneling coupling into and out of the localized state. Indeed, we numerically find identical localization behavior when directly probing the decay of eigenstates away from their maximum (not shown): ``square'' states decay exponentially away from their center region (of width $\approx 50\,\mathrm{nm}$) due to edge roughness.

At finite magnetic fields of $T\approx 1\,\mathrm{T}$, the localization length slightly increases as shown in Fig.~\ref{Figure4}e (purple curve). As soon as twice the magnetic length $\lambda_B\approx 25\,\mathrm{nm}/\sqrt{B/\mathrm{T}}$ is smaller than the ribbon width, $L_\mathrm{loc}$ abruptly increases, in good agreement with the experimental findings in  Fig.~\ref{Figure2}c.

\color{black}

In summary, we have presented electronic transport experiments of a special graphene nanoribbon geometry with added graphene wings at each side of the ribbon. While transport close to the charge neutrality point is still governed by Coulomb blockade, the details change strongly. We show that even the outer parts of the wings contribute to transport. From transport measurements in perpendicular magnetic field, we extract a typical localization area being smaller than the wing size. Previous experiments have observed states localized along the edge.~\cite{Bischoff2014} Such states are again found by the tight-binding calculations performed for the ribbon with wings devices. The comparison of the ribbons with wings with the long control ribbons shows however, that a picture where transport is only governed by charges localized along the edge is insufficient to explain the measurements. Instead we propose a picture where the different localized states are tunneling coupled. In such a picture it is meaningless to distinguish transport along the edge or over the bulk. This picture is compatible with both experiment and tight binding simulations. Notably, the simulations manage to reproduce both the typical localization length extracted from the experiment as well as the experimental behaviour in a magnetic field.

\section*{Acknowledgements}
We thank P. Simonet, A. Varlet, H. Overweg and J. Burgdörfer for helpful discussions. Financial support by the National Center of Competence in Research on “Quantum Science and Technology“ (NCCR QSIT) funded by the Swiss National Science Foundation is gratefully acknowledged. F. L. acknowledges support by the Austrian Science Foundation FWF, SFB-F41 VICOM. Calculations were performed on the Vienna Scientific Cluster.


%
%

\end{document}